\documentclass{aipproc}

\layoutstyle{8x11double}

\SetInternalRegister\hbadness{8000} 

\begin{document}

\title{The E-peak distribution of the GRBs detected by HETE FREGATE instrument}


\author{C. Barraud}{
  address={Centre d'Etude Spatiale des Rayonnements-CNRS/UPS- 9 Av du Colonel Roche,\\BP 4346-31 028 Toulouse Cedex 4,FRANCE},
  email={barraud@cesr.fr}}

\iftrue
\author{J. L. Atteia}{
  address={Centre d'Etude Spatiale des Rayonnements-CNRS/UPS- 9 Av du Colonel Roche,\\BP 4346-31 028 Toulouse Cedex 4,FRANCE},
  email={atteia@cesr.fr}}
\author{J. F. Olive}{
  address={Centre d'Etude Spatiale des Rayonnements-CNRS/UPS- 9 Av du Colonel Roche,\\BP 4346-31 028 Toulouse Cedex 4,FRANCE},
  email={olive@cesr.fr}}
\author{J. P. Dezalay}{
  address={Centre d'Etude Spatiale des Rayonnements-CNRS/UPS- 9 Av du Colonel Roche,\\BP 4346-31 028 Toulouse Cedex 4,FRANCE},
  email={dezalay@cesr.fr}}
\fi

\copyrightyear  {2001}

\begin{abstract}
The FREGATE gamma ray detector of HETE-2 is sensitive to photons between 6 and 400 keV. This sensitivity range, extended towards low energies, allows us to
explore the emission of GRBs in hard X-rays. We fit the spectra of 23 GRBs with Band's spectral function in order to derive the distribution of their peak
energies (E-peak). This distribution is then compared with the E-peak distributions measured by BATSE and GINGA.
\end{abstract}

\date{\today}

\maketitle

\section{Introduction}

We present here a preliminary analysis of the spectral distribution of FREGATE's GRBs. FREGATE is the gamma-ray detector of HETE-2 (see \cite{Atteia}
for a description of FREGATE). During its first year of operation (HETE has been launched on october 9th 2000), FREGATE has detected 37 GRB candidates:
for 13 of them, a position has been determined, 13 have no localization but are in the field of view and the remaining 11 are out of the field of view.

In this preliminary study, we focus on the distribution of E-peak which is the energy at which the $\nu F_{\nu}$ spectrum reaches a maximum. To obtain the E-peak of
a GRB, we fit its energy spectrum with a simple model (the so called Band's function) using the Xspec software.

In addition, this analysis provides the fluences of the GRBs detected within the field of view of FREGATE.

\section{The E-peak distribution}

\subsection{Spectral fitting}
The Xspec software requires 3 input components: the \emph{observed spectrum} is obtained by subtracting the background from the signal. It is coded on 128 channels
 in energy for the four detectors of FREGATE. The \emph{instrumental response} is computed from a Monte Carlo simulation of the instrument, it has been extensively
 tested with ground calibrations and inflight calibrations with the Crab nebula \cite{Olive}. The \emph{model spectrum} is the model GRBM from Xspec, based on
 Band's function \cite{Band:1993}:

\begin{tabular}{l l}
$N(E)=A.E^{\alpha}.e^{-E/E_o}$ & for $E<(\alpha - \beta)E_o$\\
$N(E)=B.E^{\beta}$& otherwise.
\end{tabular}

We can then calculate the E-peak energy ($E_p$), the energy at which the $\nu F_{\nu}$ spectrum has a maximum.\\ $E_p$ is given by:

\begin{equation}
E_p=E_o(2+\alpha)
\end{equation}

\begin{figure}[htbp]
\centering
\includegraphics[height=7.5 cm, angle=-90]{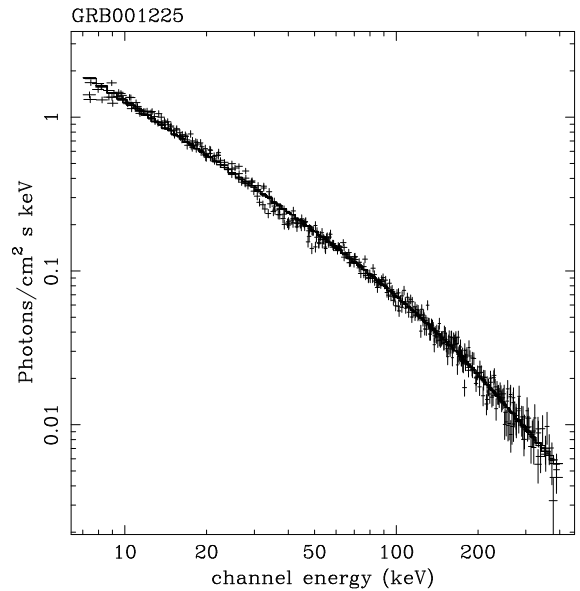}
\caption{Spectrum of GRB001225 fitted with Band's function. $\alpha = -1.15, \beta = -1.90, E_o = 277 keV, E_p = 235 keV$}
\label{GRB001225}
\end{figure}

\begin{figure}[htbp]
\centering
\includegraphics[height=7.5 cm,angle=-90]{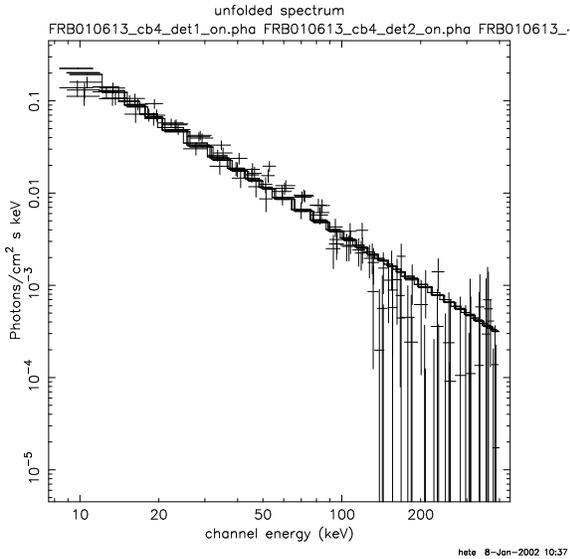}
\caption{Spectrum of GRB010613 fitted with a power law. $\alpha = -1.78$}
\label{GRB010613}
\end{figure}

$E_p$ is defined when: $\alpha > -2$ and $\beta<-2$. \\As an example the figure \ref{GRB001225} shows the spectrum of GRB001225 which is well fitted by the
Band's model. In comparison, the figure \ref{GRB010613} shows a spectrum which can be fitted by a simple power law.

The table \ref{sursauts} lists the spectral parameters of 23 GRB candidates which occured within the FOV of FREGATE. Most of these events were confirmed by
other GRB detectors on ULYSSES, KONUS, SAX-GRBM, RXT-ASM.

\subsection{The E-peak distribution}

\begin{figure}[htbp]
\centering
\includegraphics[width=7.5 cm]{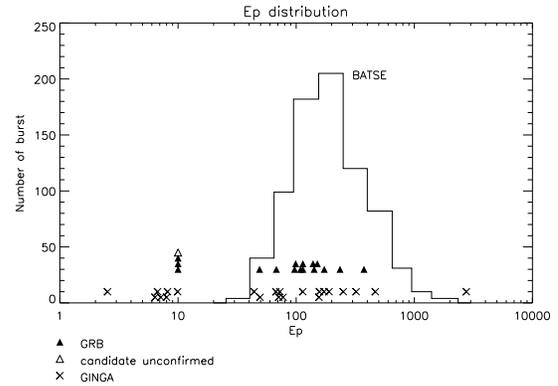}
\caption{The E-peak distribution of the GRBs detected by FREGATE (triangles), GINGA (crosses) and BATSE.}
\label{distribution}
\end{figure}

The figure \ref{distribution} shows the E-peak distributions mesured with the GRBs detected by FREGATE compared to BATSE and GINGA.
The value of $E_P$ for FREGATE GRBs is obtained with Xspec, the distribution of BATSE comes from \cite{Brainerd} and the GINGA's distribution was
calculated using \cite{Strohmayer:1998}.

Some GRBs can be fitted by a simple power law (in that case $\beta$ is arbitrarely fixed at -10). When this is the case, if $\alpha \ge -2$, we consider that the data
do not allow us to derive $E_p$ and we provide no value (see GRB010613). if $\alpha \le -2$, the $\nu F_{\nu}$ spectrum is steadily decreasing in the energy
range of FREGATE and we arbitrarily set $E_p$ to 10 keV.

The spectral fitting of FREGATE GRBs also allows us to compare the fluences in various energy ranges.

\subsection{The fluence}

\begin{figure}[htbp]
\centering
\includegraphics[width=7.5 cm]{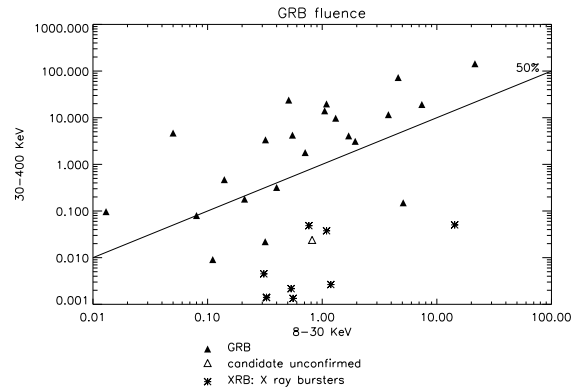}
\caption{The fluence in 30-400 keV vs the fluence in 8-30 keV for 23 GRBs detected by FREGATE. The XRFs are placed under the line $50\%$.}
\label{fluence}
\end{figure}

The figure \ref{fluence} shows the fluence in the range 30-400 keV versus the fluence in the range 8-30 keV. The line $50 \%$ represente the place where there is as
much fluence in the range 8-30 keV as in the range 30-400 keV. These two fluences are well correlated for confirmed GRBs (with $80\%$ of the fluence in the range
30-400 keV on average).

A few events are very soft, having most of their fluence below 30 keV (they are located below the line $50\%$). The nature of these events is still unclear: genuine GRBs,
very soft GRBs... see also \cite{heise}. According to a suggestion by J.Heise (GCN 1138), we call them X-Ray Flashes (XRFs).

For comparison, a few typical X-ray bursts have been added to this figure.

\begin{table}[!t]
\begin{tabular}{|c|c|c|c|c|c|c|c|}
\hline
Sursauts & angle & fluence& fluence& $ \alpha $ & $\beta$ & $E_o$ & $E_p$\\
&&8-400 keV&30-400keV&&&keV&keV \\ \hline
GRB001102 \footnote{This burst was in the limit of the field of view of FREGATE} & $70^o$ &77.33&72.33& 0.14 & -2.86 &58 &142\\ \hline
GRB001225 &$37^o$&165.12&143.63&-1.15 &-1.90 &277& 235\\ \hline
GRB010126 &$10^o$&2.50&1.79&-.92 &-3.46 &90&97 \\ \hline
GRB010213&$14^o$&.12&.0091 & -1.50 &-10&12&\\ \hline
GRB010225 &$23^o$&.39&.18&-1.27 &-10 &37 &\\ \hline
GRB010326a &$60^o$&15.10&14.05&-0.13 &-3.12 &81&151 \\ \hline
GRB010326b &$17^o$&.16&.08&-2.13 &-10 &50 & 10\\ \hline
GRB010612 & $13^o$&4.78&4.23&-0.90 &-10   & 189&\\ \hline
GRB010613 & $38^o$&26.58&19.18& -1.15& -10&80&\\ \hline
GRB010629 &$28^o$& 5.06&3.12&-1.41 &-3.50 &83&49\\ \hline
GRB010921 &$44^o$&15.41&11.63 &-1.42 &-2.35 &170&99\\ \hline
GRB010923 &$60^o$&5.78&4.08&-0.57 &-3.63 &80&114\\ \hline
GRB010928&$24^o$&20.86&19.77&-0.64&-1.82&276&375\\ \hline \hline
GRB001115a & $30^o-150^o$&24.41&23.90& 0.53 &-1.55 &55&139\\ \hline
GRB001115b & &11.07&9.76& -0.91&-1.42&105&114\\ \hline
GRB010428 &$4^o-82^o$& .34&.022&-2.06 &-10 &12&10\\ \hline
GRB010609 &$48^o-195^o$&5.24&.15 &-3.14 &-10 &11&10\\ \hline
GRB010621\footnote{this events have been detected only by FREGATE} & &.84&.024&-2.31 &-10 &10&10\\ \hline
GRB010706 & &.11&.097&0.29 &-10 &58&\\ \hline
GRB010827a &&.61&.47 &-.79 &-2.44 &56&68\\ \hline
GRB010828 & $46^o-113^o$&.72&.32&-0.96 &-10 &176&\\ \hline
GRB010903 & $53^o-106^o$&3.68&3.36&0.10 &-3.53 &52&109\\ \hline
GRB010917 & &4.74&4.69&0.99&-1.69 &58&173\\ \hline
\end{tabular}
\caption{\label{sursauts}GRB candidates detected by FREGATE}
\label{sursauts}
\end{table}

\section{conclusion}
A preliminary study of 23 GRB candidates detected by FREGATE within its FOV shows the existence of a few soft events, XRFs (with $E_p < 15$ keV and most of the
fluence below 30 keV).These XRFs can be compared to the detection of such soft events by GINGA and the WFC on Beppo SAX.

 In its second year of observations, HETE-2 will provide fast localizations of several new XRFs, an approach which may shed a new light on the nature of these
 enigmatic sources.


\bibliography{article}

\end{document}